\documentstyle[agupp,graphicx]{article}  % double colonne, simple interligne
\authoraddr{Agn\`es Helmstetter, Lamont-Doherty Earth Observatory,
61 Rte 9W, Palisades, NY 10964 (e-mail: agnes@ldeo.columbia.edu)}
\authoraddr{Bruce Shaw, Lamont-Doherty Earth Observatory,
61 Rte 9W, Palisades, NY 10964 (e-mail: shaw@ldeo.columbia.edu)}

\newcommand \be{\begin{equation}}
\newcommand \ba{\begin{eqnarray}}
\newcommand \ee{\end{equation}}
\newcommand \ea{\end{eqnarray}}
\lefthead{Helmstetter and Shaw }
\righthead{Stress heterogeneity and aftershock rate}

\begin{document}
\setkeys{Gin}{draft=false} % otherwise does not plot figures in draft mode...
\def\today{\ifcase\month\or
 January\or February\or March\or April\or May\or June\or
 July\or August\or September\or October\or November\or December\fi
 \space\number\day, \number\year}

\title{Relation between stress heterogeneity and aftershock rate in the rate-and-state model}

%\hfil File created \today

\author{Agn\`es Helmstetter and Bruce E. Shaw}
\affil{Lamont-Doherty Earth Observatory, Columbia University, New York}

\begin{abstract}
We estimate the rate of aftershocks triggered by a heterogeneous
stress change, using the rate-and-state model of {\it Dieterich} [1994].
We show that an exponential stress  distribution $P_{\tau}(\tau) \sim \exp(-\tau/\tau_0)$ 
gives an Omori law decay of aftershocks with time $\sim1/t^p$, with an exponent 
$p=1-A\sigma_n/\tau_0$, where $A$ is a parameter of the rate-and-state friction law, 
and $\sigma_n$ the normal stress.
Omori exponent $p$ thus decreases if the stress "heterogeneity" $\tau_0$ decreases. 
We also invert the stress distribution  $P_{\tau}(\tau)$ 
from the seismicity rate $R(t)$, assuming that the stress does not change with time.
We apply this method to a synthetic stress map, using the (modified) 
scale invariant "$k^2$" slip model [{\it Herrero and Bernard}, 1994].
We generate synthetic aftershock catalogs from this stress change.
The seismicity rate on the rupture area shows a huge increase at short times, 
even if the stress decreases on average. 
Aftershocks are clustered in the regions of low slip, but the spatial 
distribution is more diffuse than for a simple slip dislocation.
Because the stress field is very heterogeneous, there are many patches of 
positive stress changes everywhere on the fault.
This stochastic slip model gives a Gaussian stress distribution, 
but nevertheless produces an aftershock rate which is very close to Omori's law, 
with an effective  $p\leq 1$, which increases slowly with time. 
We obtain a good estimation of the stress distribution for realistic catalogs, when we constrain 
 the shape of the distribution. 
 However, there are probably other factors which 
also affect the temporal decay of aftershocks with time.
 In particular, heterogeneity of $A\sigma_n$ can also 
 modify the parameters $p$ and $c$ of Omori's law.
 Finally, we show that stress shadows are very difficult to observe in a
heterogeneous stress context.  

\end{abstract}

\begin{article}

\section{Introduction}

Much progress has been made in describing earthquake behavior
based on the predictions of rate-and-state friction. 
The rate-and-state  model explains the $1/t$ decay  
of aftershock rate as a function of the time $t$ since the mainshock (Omori's law)
independent of the mainshock magnitude,
the scaling of aftershock duration with stressing rate, 
the slow diffusion of aftershocks with time [{\it Dieterich}, 1994].  
This success led several authors to provide time-dependent 
earthquake probabilities using this model [{\it Toda et al.}, 1998; 2003; 2005]. 
Many other physical mechanisms have been proposed to explain 
Omori law, such as sub-critical crack 
growth [{\it Das and Scholz}, 1981; {\it Shaw}, 1993], viscous relaxation 
[{\it Mikumo and Miyatake}, 1979], static fatigue 
[{\it Scholz}, 1968; {\it Narteau et al.}, 2002], 
postseismic slip [{\it Schaff et al.}, 1998],
or pore fluid flow [{\it Nur and Booker}, 1972].
The rate-and-state model of {\it Dieterich} [1994] is probably the best candidate, 
however,
because it only relies on a rate-and-state dependent
friction law observed in laboratory experiments.

At the same time, a number of fundamental puzzles remain.
One of the most striking is the abundance of aftershocks on the
rupture surface, where indeed most aftershocks occur.
This is in stark contrast with simple pictures of the rupture process, 
which suggest stress should have decreased on the rupture surface
and there should therefore be a dearth of aftershocks there.
A second fundamental puzzle concerns the time dependence of aftershocks.
Here, subtle but significant deviations from the pure Omori law 
inverse time decay of the rate of aftershocks is seen in averages
of aftershock rates [{\it Helmstetter et al.}, 2005].  While
{\it Dieterich} [1994] explained this as a consequence of the spatial
dependence of stress as it decreases away from the fault,
or as a change of stressing rate with time,
such mechanisms do not seem to properly explain the 
aftershocks occurring on the rupture area.
Thus, both the spatial and temporal distribution of the 
majority of aftershocks have yet to be fully explained.
Here, we show how an extension of the rate-and-state formulation, 
which takes as its foundation a heterogeneous stress field,
can explain these observations.  We then use this
model to estimate 
stress heterogeneity from aftershock rates.

Our work builds off of the pioneering work of {\it Dieterich} [1994], 
who derived a relation between seismicity rate and stress history, 
for a population of faults obeying rate-and-state friction.
For a uniform positive stress step (e.g., a mainshock), the rate-and-state 
model gives an Omori law decay of the seismicity rate $R(t) \sim t^{-p}$ 
with $p=1$ for intermediate times. At very short times, smaller than a characteristic 
time $c$, which depends on the stress change, the seismicity rate is constant. 
{\it Dieterich} [1994] also computed the aftershock rate for a dislocation, 
with a uniform stress decrease on the rupture area, and a positive stress change outside 
the rupture, decaying as $\tau \sim 1/\sqrt{r}$  in the near field, and $\tau \sim 1/r^3$ 
in the far field for $r\gg L$. As distance from the fault increases, the 
characteristic time $c$ (typical time between mainshock and aftershocks) increases. 
Integrating over the fault, the seismicity rate approximately obeys Omori law 
$R(t) \sim 1/t^p$, with an apparent exponent $p<1$.
 
{\it Dieterich et al.} [2000, 2003] used the rate-and-state model of seismicity 
to invert stress history from seismicity rate, and apply this method to Hawaii seismicity.
They discretize the space, with a grid size of about 1 km, and assume that the stress 
is uniform in each cell. This method then gives the stress history in each cell.
The assumption that the stress is uniform at scales of a few  km is reasonable 
for the stress change induced by a dyke intrusion, as in [{\it Dieterich}, 2000, 2003],
or for the coseismic stress change induced by a large earthquake in the far field.
However, the coseismic stress change on the mainshock fault plane, where most 
aftershocks occur, is probably very heterogeneous at all scales 
[{\it Herrero and Bernard}, 1994].

%Here we try to estimate the stress distribution on the fault plane 
%from the aftershock rate, using the rate-and-state model. 
%We assume that the stress changes instantaneously after the mainshock, and we 
%neglect the relaxation of stress on the fault due to aseismic slip or viscous relaxation.
%We also neglect the stress change, and seismicity rate change, induced by aftershocks. 

In this paper, we investigate how heterogeneity of the Coulomb stress change 
and of the normal stress modifies the temporal decay 
of aftershocks with time, both on the  fault and off-the fault. 
We assume that the stress changes instantaneously after the mainshock, and we 
neglect the relaxation of stress on the fault due to aseismic slip or viscous relaxation.
We also neglect the stress change, and seismicity rate change, induced by aftershocks. 
We then try to invert for the stress distribution on the fault plane 
from the aftershock rate, using the rate-and-state model, and assuming the main source 
 of heterogeneity is the coseismic stress change. 

\section{Relation between stress distribution and seismicity rate}

{\it Dieterich} [1994] derives a differential equation which gives the
seismicity rate $R(t,\tau)$ as a function of the stress history $\tau(t)$.
His model assumes an infinite population of faults which obeys rate
and state friction, with the same properties for all faults.

The state variable $\gamma$ is related to the stress $\tau$ by
\be
\partial \gamma={1 \over A \sigma_n} [\partial t -\gamma \partial \tau]~,
\label{dg}
\ee
where  $\tau$  is the "modified" Coulomb stress change [{\it Dieterich et al.}, 2000], 
and $\sigma_n$ is the normal stress.
The state variable $\gamma$ is a function of the seismicity rate $R(t,\tau)$
\be
R(t,\tau)= {R_r \over \gamma(t,\tau) \dot{\tau_r}}~,
\label{gamma}
\ee
where $R_r$ is the steady state seismicity rate at the reference
stressing rate $\dot{\tau_r}$.
From laboratory experiments, coefficient $A$ generally has values between 
0.005 and 0.02, for various temperature and pressure conditions [{\it Dieterich}, 1994].

\begin{center}
\begin{figure}
\includegraphics[width=0.5\textwidth]{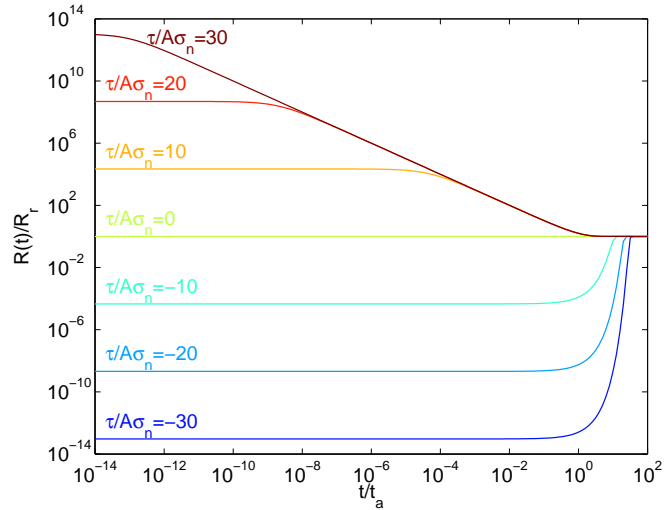}
\caption{\label{figRs} Seismicity rate $R(t,\tau)$ 
(normalized by the reference seismicity rate $R_r$) 
as a function of time (normalized by the aftershock duration $t_a$), given by the 
rate-and-state model with a uniform stress step (\ref{R}), 
for different values of the stress change, 
ranging from $\tau/A\sigma_n=-30$ (bottom) to  $\tau/A\sigma_n=30$ (top). }
\end{figure}
\end{center}

{\it Dieterich} [1994] used expression (\ref{dg}) to derive the seismicity rate
$R(t,\tau)$ triggered by a single stress step $\tau$. 
We assume that stress rate after the stress step is constant 
$d\tau /dt = \dot{\tau_r}$, and that the seismicity rate before the 
mainshock is equal to the reference seismicity rate $R_r$.
Using  (\ref{dg}), the seismicity rate following the stress step is
\be
R(t,\tau)={R_r \over \left( e^{ - \tau / A \sigma_n}-1
\right)~e^{-t/t_a}+1} \,, 
\label{R}
\ee
where $t_a$ is the duration of the aftershock sequence 
\be
t_a = {A \sigma_n \over \dot{\tau_r}} \,.
\label{ta}
\ee
This relation (\ref{R}) is illustrated in Figure~\ref{figRs} for 
different values of the stress change. 
For each positive stress value, the seismicity rate is constant for
 $t\ll t_a e^{-\tau/A\sigma_n}$,
and then decreases with time for $ t_a e^{-\tau/A\sigma_n} \ll t \ll t_a$ according to 
Omori law with an exponent $p=1$. For a negative stress change, the seismicity rate 
decreases after the mainshock. In both cases, the seismicity rate recovers its reference 
value $R=R_r$ for $t\gg t_a$. 
The goal of this work is to extract the stress distribution from the seismicity rate.
This is a difficult problem, because, as shown in Figure~\ref{figRs}, the 
seismicity rate does not depend on the stress change 
over a relatively large time interval.

For a heterogeneous stress field $\tau(\vec r)$, with a distribution 
(probability density function) $P_{\tau}(\tau)$, the seismicity rate integrated over space is
\ba
R(t) &=& \int \, R(t,\tau(\vec r)) \, d\vec r  \label{Rt1} \\
     & = & \int \limits_{-\infty}^{\infty} \, R(t,\tau) 
     \, P_{\tau}(\tau) \, d\tau \label{Rt2} \\
     & = & \int \limits_{0}^{\infty} \, R(t,c) \, P_c(c) \, dc \label{Rt3}
\ea
where $c=t_a \, e^{-\tau/A\sigma_n}$ is a characteristic time of the aftershock rate, 
such that
$R(t,c) \sim 1/c $ for $t\ll c$ and $R(t) \sim 1/t$ for $c \ll t \ll t_a$.

Equation (\ref{Rt3}) is a Fredholm integral equation of the first kind. 
It has, at most, one solution [{\it Riele}, 1985].
Equation (\ref{Rt3})  has a simple approximate solution in the case when 
the stress change has an exponential distribution  
\be
P_{\tau}(\tau) \sim e^{- \tau /\tau_0} \, ,
\label{ps}
\ee 
 where $\tau_0$ is a positive scaling stress 
parameter, which characterizes the width of the stress distribution.
This corresponds to a power-law distribution of 
%bs added quotes to corner times so as not to confuse 
% times with multiplied by
``corner times'' $c$
\be
P(c) = P_{\tau}(\tau) \, {d\tau \over dc} = c^{-1+ A\sigma_n /\tau_0} \,.
\label{pc}
\ee
We also consider an approximate expression for the seismicity rate (\ref{R}) valid 
for short times $t\ll t_a$ 
\be
R(t,\tau) \approx { R_r \over  e^{ -\tau /A\sigma_n } + t/t_a} = { R_r t_a \over  c + t} \,. 
\label{R3}
\ee
Substituting (\ref{pc}) and (\ref{R3}) in (\ref{Rt3}),  we  get
\be
% R(t) \sim {1 \over t^{1- 1/\tau_0}} \, , 
%% error, missing A \sigma_n , and added the integral below
R(t)  = \int \limits_{0}^{\infty} \, 	{ R_r t_a  c^{-1+ A\sigma_n /\tau_0} \over  c + t} 	 \, dc 
\sim {1 \over t^{1- A \sigma_n/\tau_0}} \, , 
\label{Rte2}
\ee    
Expression (\ref{Rte2}) corresponds to Omori law with an exponent 
\be
p=1- { A \sigma_n \over \tau_0} \,.
\label{p}
\ee
%This result shows that, in order to observe a pure Omori law at short times 
%with an exponent $p<1$, the stress field must have an exponential distribution 
%in the tail, for large positive stress. 
%%
Because equation (\ref{Rt2})  has at most one solution, the exponential stress distribution is 
the only distribution which produces a pure Omori law decay for $t\ll t_a$, 
without any cut-off or crossover at short times.
However, other distributions, e.g., a Gaussian, produce aftershock rate 
that is very close to Omori's law, over a very large time range.
The stress distribution for small or negative values is not constrained by
the seismicity rate at short times $t\ll t_a$,  so deviations from an exponential for 
negative stresses does not produce deviations from Omori law at short times.

\begin{center}
\begin{figure}
\includegraphics[width=0.5\textwidth]{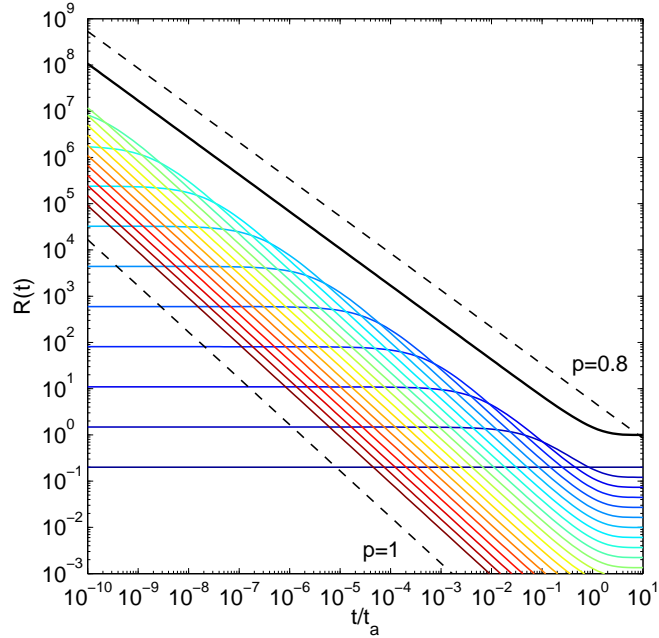}
% figRps.m
\caption{\label{figRtPexp} 
The thin colored lines are the seismicity rate
$R(t,\tau)$ for a uniform stress change $\tau$, ranging from 
$\tau=0$ (blue flat curve) to $\tau=50$ MPa (red curve),
weighted by the probability $P_{\tau}(\tau)$, using $A\sigma_n=1$ MPa.
The stress distribution is given by $P_{\tau}(\tau) \sim \exp(-\tau/5)$ with $\tau>0$.
The solid black line is the total seismicity rate $R(t) = \int_0^\infty R(t,\tau) ~ P_{\tau}(\tau) ~d\tau$. 
The superposition of curves $R(t,\tau)$ with 
a power-law distribution of crossover times $c =t_a \exp(-\tau /A\sigma_n)$ 
gives rise to a power law decay of $R(t)$ with an exponent $p\approx 0.8$. 
The dashed lines are Omori laws with $p=1$ (bottom) and $p=0.8$ (top).}
\end{figure}
\end{center}

Expression (\ref{p}) shows that Omori exponent depends on stress heterogeneity. 
The parameter $\tau_0$ represents the width of the stress distribution for $\tau>0$.
The more heterogeneous the stress is (larger $\tau_0$), the larger $p$ is (closer to 1). 
Figure~\ref{figRtPexp} illustrates how the rate-and-state model with a heterogeneous 
stress distribution produces a power-law decay with an exponent $p<1$.

{\it Helmstetter et al.} [2005] found that, for stacked aftershock sequences in Southern 
California, Omori exponent is close to 0.9, for times ranging between a minute 
(but possibly even less) and one year, and for mainshock magnitudes between 2 and 7.5.
This suggests that the stress distribution is close to exponential in the tail, 
with a characteristic stress  $\tau_0 \approx 10 A\sigma_n$.
Assuming that $A=0.01$ (as measured in laboratory friction experiments 
[{\it Dieterich}, 1994]) and $\sigma_n=100$ MPa (corresponding to the lithostatic 
pressure at a depth of about 5 km), this gives $A\sigma_n=1$ MPa and $\tau_0=10$ MPa, 
a value larger than the typical stress drop $\sigma_0=3$ MPa [{\it Ide and Beroza}, 2001], 
but of the same order of magnitude.
However, a few studies tried to estimate $A\sigma_n$ directly from earthquake catalogs, 
and obtained values smaller than the ones derived from the laboratory value of $A$.
{\it Dieterich} [1994] found $A\sigma_n=\sigma_0/20$, from the relation between aftershock duration 
and the recurrence time (assuming characteristic earthquakes). This gives  $A\sigma_n=0.15$ MPa
 assuming 
%bs deleted: character 
a stress drop of 3 MPa. 
{\it Cochran et al.} [2004] used the rate-and-state model to model tidal triggering of earthquakes,
and obtained a prefered value of $A\sigma_n=0.064$  MPa, and an acceptable range $
0.048< A\sigma_n< 0.11$ MPa. 

The rate-and-state model with a uniform stress step (\ref{R}) cannot explain an Omori law decay with $p>1$. 
Equation (\ref{Rt3}) does not have a solution with $P_{\tau}(\tau)>0$ and $t\ll t_a$ in this case. 
Some aftershock sequences however have an Omori exponent larger than one. 
The only solution in order to obtain a $p$-value larger than one 
in the rate-and-state model is
to have a variation of stress with time, which may be due to postseismic slip 
or viscous relaxation, although these explanations
involve relatively large stress changes with time [{\it Dieterich}, 1994]. 
Other explanations for Omori's law do allow for larger $p$-values
[{\it Mikumo and Miayatake}, 1979; {\it Shaw}, 1993; {\it Narteau et al.}, 2002].

%%%%%%%%%%%%%%%%%%%%%%%%%%%%%%%%%%%%%%%%%%%%%%%%%%%%%%%
\section{Estimating the stress distribution from aftershock rate \label{secinv}}

We have shown above that, according to the rate-and-state model,
the Omori exponent provides some information on the stress heterogeneity 
(but only if $p<1$). Furthermore, we can (in theory) obtain 
the complete stress distribution (in the region where we measure the seismicity rate) 
from the temporal evolution of the seismicity rate.  
Expression (\ref{Rt2}) indeed provides a method for estimating the 
full distribution $P_{\tau}(\tau)$, provided we observe 
the seismicity rate $R(t)$ over a wide enough time interval.

We first discretize the integration over stress and times, 
using a linear sampling for stress, and a logarithmic sampling for times, 
using the same number $N$ of points.
Equation (\ref{Rt2}) is then similar to the system of $N$ linear equations
\be
R(t_j) = \sum \limits_{i=1}^N \, R(t_j,\tau_i)\, P_{\tau}(\tau_i) \, (\tau_{i+1} -\tau_{i}) 
\label{Mp}
\ee
We divide both sides of equation (\ref{Mp}) 
by $R(t_j)$ to stabilize the problem. Equation (\ref{Mp}) thus becomes 
\be
1 = \sum \limits_{i=1}^N \, { R(t_j,\tau_i) \over R(t)} \,
 P_{\tau}(\tau_i) \, (\tau_{i+1} -\tau_{i}) = M \times P
\label{Mp2}
\ee
where $M$ is a $N\times N$ matrix $M(i,j)=  (\tau_{i+1} -\tau_{i}) \, R(t_j,\tau_i)/R(t_j)$ 
and the vector $P$ is the stress distribution at points $\tau_1,...,\tau_N$.

The inversion of the stress distribution from (\ref{Mp2})
is an ill-posed problem, i.e., the solution is very sensitive to noise. 
We thus use the regularization method of [{\it Riele}, 1985]. 
We introduce an additional constraint to (\ref{Mp}), minimizing either 
the first derivative  $|P'(\tau)|$, the smoothness $||P''(\tau)||$,     
or the distance between $P_{\tau}(\tau)$ and an initial guess $P_0(\tau)$.
(e.g., a Gaussian distribution).
Instead of solving directly (\ref{Mp2}), we minimize the quantity
\be
|| M \, P - 1 ||^2 + \alpha || L(P)||^2 \,,
\label{alpha}
\ee
where $\alpha>0$ is the regularization parameter, and $L$ 
is a linear operator, e.g., $L(P)=P-P_0$,  $L(P)=P'$ 
(first derivative), or  $L(P)=P''$ (second derivative). 
We also impose that the stress distribution is positive.
We thus search for the positive vector $P$ that minimizes equation 
(\ref{alpha}), using the non-linear least-square fitting program 
given by {\it Lawson and Hanson} [1974]. 

In practice, the estimation of $P_{\tau}(\tau)$ for large $\tau$ is limited by 
the minimum time $t_{\rm min}$ at which we can reliably estimate the seismicity rate.
The largest stress we can resolve is of the order of $\tau_{\rm max} = 
 - A\sigma_n\log (t_{\rm min}/t_a)$. 
 Practically, this time $t_{\rm min}$ may be as low as a few
 seconds, if we correct from catalog incompleteness shortly after the mainshock 
 [{\it Vidale et al.}, 2004].
For negative stress, we are limited by the maximum time $t_{\rm max}$ 
after the mainshock, and by our assumptions that secondary aftershocks are negligible, 
and that the stress does not change with time (e.g., neglecting post-seismic relaxation).
In order to resolve $P_{\tau}(\tau)$  for negative values, we need to know the seismicity rate 
for times larger than the aftershock duration $t_a$ (i.e., usually 
at least a few years).
Indeed, the seismicity rate after a stress decrease is close to zero for $t\ll t_a$, 
so that the measure of $R(t)$ for $t\ll t_a$ does not provide any information on 
$P_{\tau}(\tau)$ for $\tau < 0$.

\section{Application of the method to a stochastic slip model}

\subsection{Stochastic $k^2$ slip model}

We have tested the rate-and-state model on a realistic synthetic slip pattern.
{\it Herrero and Bernard} [1994] proposed a kinematic, self-similar model 
of earthquakes.
They assumed that the slip distribution at small scales, compared to the rupture 
length $L$, does not depend on $L$. This led to a slip power-spectrum for high 
wave-number equal to
\be
 u (k) = C { \sigma_0 \over \mu} {L\over k^2}
\;\;\; \mbox{for} \; k > 1/L \, ,
\label{uk2}
\ee
where $\sigma_0$ is the stress drop (typically 3 MPa), $\mu$ 
is the rigidity (typically 3300  MPa in the lower crust), and $C$ is a 
shape factor close to 1. 
For wavelengths larger than the rupture length $L$, the 
power spectrum is constant 
 \be
 u (k) = C {\sigma_0 \over \mu} L^3 
\;\;\; \mbox{for} \; k < 1/L \, .
\label{uk22}
\ee 
This model (\ref{uk2}) reproduces the $1/f^2$ power-spectrum 
of seismograms for large frequencies [{\it Herrero and Bernard}, 1994].

\subsection{Shear stress change and seismicity rate on the fault}
We have used the $k^2$ model to generate a synthetic slip pattern, 
and compute the shear stress change on the fault from the slip 
[{\it Andrews}, 1980; {\it Ripperger and Mai}, 2004].
Note that the seismicity rate given by (\ref{R}) depends on the Coulomb stress
change, which is equal to the shear stress change on the fault  because
the normal stress change on a planar fault is zero. 
If we analyze off-fault aftershocks or complex rupture
%bs added word below
geometries, 
we would have to consider changes in normal stress as well.

We have modified the $k^2$ model in order to have a finite 
standard deviation of the stress distribution.
The $k^2$ model (\ref{uk2}) produces a shear stress change 
with a power spectrum $\tau(k) \sim k^{-1}$ for large $k$, because
 the stress is approximately the derivative of the slip.
 As a consequence, the shear stress change for the $k^2$ model  
is extremely heterogeneous, with an infinite standard deviation.
The exponent $n=2$ in  the $k^2$ model (\ref{uk2}) is thus 
a minimum physical value for the slip power-spectrum [{\it Herrero and 
Bernard}, 1994]. Using $ u(k) \sim k^{-2}$ produces a shear stress change 
with a standard deviation which diverges logarithmically as the
maximum wavenumber increases. Thus Omori $p$-value for this slip model 
tends to 1 as the grid resolution increases.
We have thus replaced the exponent $n=2$ in (\ref{uk2})  
by $n=2.3$, and smoothed the crossover at $k = 1/L$, using
\be
 u(k) =  C {\sigma_0 \over \mu} { L^3 \over (kL +1)^{n}}
\, .
\label{uk23}
\ee
 
We have computed the stress change on the fault from this synthetic 
slip model, for a fault of $50\times50$ km, with a resolution $dx=0.1$ km, 
and a stress drop $\sigma_0= 3$ MPa (i.e., the average stress change on the 
fault is $-3$ MPa).  The maps of the slip and stress on the fault are 
 shown in Figure~\ref{figUS}. The stress field has large variations, 
 from about -90 to 90 MPa, due to slip variability.
We did not constrain the slip to be positive. 
This could be done by changing the phase of the lowest mode, and tapering the slip
close to the edges, so that the maximum slip is at the center [{\it Herrero and Bernard}, 1994]. 
Doing so introduces small deviations of the stress distribution from a Gaussian distribution for 
$\tau \approx 0$, but does not introduce significant changes on the seismicity rate .

We have then estimated the seismicity rate on the fault predicted by the rate-and-state 
model, by integrating numerically  (\ref{Rt1}) using the observed
 stress map, and  $A\sigma_n=1$ MPa.  
  While the stress on average decreases on the fault, the seismicity rate  
 shows a huge increase after the mainshock (by a factor $10^{10}$, but, 
 of course, the seismicity rate  at short times, smaller than the duration 
 of the earthquake, has no physical sense) (see Figure~\ref{figRt1}). 
 It then decays with time approximately according to 
 Omori law, with an apparent exponent $p=0.93$. At large times $t\approx t_a$, 
 the seismicity rate decreases below its reference rate due to the negative stress 
 values. 

 {\it Marsan} [2006] reached similar conclusions, using the same model:
 the main effect of stress heterogeneity on the  fault is to produce a short term 
 increase of the seismicity rate, and to delay the  seismic quiescence on the fault   by months to years.

 \subsection{Synthetic aftershock catalog}
 
 We have generated synthetic earthquake catalogs 
 according to the rate-and-state model, using the (modified) $k^2$ model 
(\ref{uk23}) to generate the stress change. 
We have simulated aftershock sequences triggered by this heterogeneous 
stress change, without including earthquakes interaction (i.e., without 
coseismic stress changes induced by aftershocks), using the method 
of {\it Dieterich et al.} [2003].
We assume a non-stationary Poisson process with an average 
seismicity rate $R(t,\tau)$ given by  (\ref{R}).
We generate aftershock in each cell independently of the other cells, assuming that the stress 
is uniform in each cell. 
We do not need to generate event magnitudes, because we do not include 
secondary aftershocks in our simulation. We consider that each aftershock does not modify 
the stress field or the seismicity rate.

In each cell, we generate events one after the other.
If the last event in the cell occurred at a time $t_i$ after  the mainshock , 
the probability that the next earthquake will occur at 
a time smaller than $t_i+dt$ is given by
\be
F(dt,t_i) = \exp \Bigl[ -\int _{t_i}^{t_i+dt} R(t',\tau) ~dt' \Bigr] 
\label{Ft}
\ee
The function $F(dt,t_i) $ increases from 0 to 1 as $dt$ ranges from 0 to $\infty$.
To determine the time $t_{i+1}=t_i+dt$ of the next event, we generate a random number 
 $z$ between 0 and 1, and we solve for $F(dt,t_i) =z$.
%Evaluating this expression and solving for $x$ numerically is nontrivial, 
%and several numerical approximations need to be made.
%%
We have generated 6 synthetic catalogs from the stress field shown in 
Figure~\ref{figUS}, using $A\sigma_n=1$ MPa or $A\sigma_n=0.1$ MPa. We used different values 
values of the reference rate $R_r$, and of time interval $t_{\rm min} - t_{\rm max}$ 
(see Table~1), in order to test how the inversion method depends on the quality of the catalog.

 \subsection{Inversion of stress history from seismicity rate}
 
\begin{center}
\begin{figure}
\includegraphics[width=0.5\textwidth]{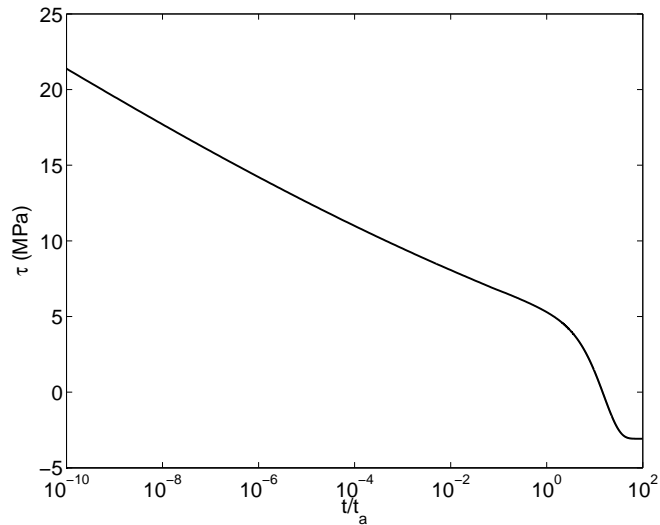}
\caption{\label{figSt} 
 Shear stress change, as a function 
of the time after the mainshock, estimated from the seismicity rate using 
{\it Dieterich et al.} [2000, 2003] method. We solved equation (\ref{dg})
 for the stress history, assuming that the stress is uniform in space but 
 changes with time.}
\end{figure}
\end{center}

 We have first applied the method of {\it Dieterich et al.} [2000, 2003] 
 on this synthetic stress field shown in  Figure~\ref{figUS}b.
{\it   Dieterich et al.} [2000, 2003] estimate the stress history $\tau(t)$ 
at any point on a grid, assuming that the stress change 
 is homogeneous in each cell, but may change with time. The stress history is obtained 
 from the seismicity rate by solving equation (\ref{dg}). 
  We wanted to apply this method on this synthetic stress model to test 
 how stress heterogeneity affects the inverted stress change. 
  The results are shown in Figure~\ref{figSt}. 
  The inverted stress change at short times is close 
 to the maximum stress change $\approx 100$ MPa, and then decreases down to a value close
 to the average stress change  $\approx -3$ MPa at large times $t>t_a$.
 Dividing the fault into smaller size cells would not improve the results very much.
 Because this slip model is self-similar, there are almost everywhere 
 some parts of the fault where the stress (and thus the seismicity rate) increases.  
 This shows that a small-scale stress heterogeneity, without any time 
dependence,  is interpreted by this method as a variation of stress with time. 
 Also, it shows that a stress decrease cannot be resolved if it is mixed with 
a stress increase, unless looking at very long times. This may explain why 
 stress shadows are so difficult to observe [{\it Felzer et al.}, 2005]. 

\subsection{Inversion of stress distribution from seismicity rate}

This test shows that variability with time is hard to distinguish from 
 small-scale heterogeneity in space based on the temporal evolution of the seismicity rate.

In order to characterize the coseismic stress change on the fault plane, 
we thus need to neglect one effect (small-scale heterogeneity) or the other 
(time variation). Our method estimates the stress distribution on the fault 
from the seismicity rate, assuming that stress does not change with time.
In theory  (if we had an infinite time interval, a huge number of aftershocks, 
no foreshocks or secondary aftershocks, and if we knew the parameters 
$R_r$, $t_a$, and $A\sigma_n$), this method provides the distribution of stress 
on the fault. If the fault is divided into smaller cells, this method 
gives a map of the average stress change in each cell, as well as its variability. 

For each synthetic catalog, we have measured the seismicity rate on the fault 
by smoothing aftershock times. We used a kernel method to 
estimate $R(t)$ from aftershocks time $t_i$, with $i=1$ to $N$, with a log-normal 
filter
\be 
R(t) = \sum_{i=1}^{N} {1 \over h t \sqrt {2 \pi}} \, \exp 
\left( - {(\log_{10} (t) - \log_{10}(t_i))^2 \over 2 h^2} \right)
\label{kern}
\ee
with a kernel width $h=0.08$.

We then used the inversion method described 
%bs no section number now, so deleted: in section \ref{secinv}
%bs and added word below
previously
to estimate the stress distribution  $P_{\tau}(\tau)$ from the seismicity rate. 
We used the regularization condition $L(P)=P'$ in (\ref{alpha}), i.e., 
minimizing the derivative of $P_{\tau}(\tau)$,
using $\alpha=10^4$ (decreasing $\alpha$ produces huge fluctuations of $P_{\tau}(\tau)$).

We have also estimated the Gaussian stress distribution that best fits 
the observed seismicity rate. We evaluate the mean $-\sigma_0$, and the standard deviation 
$\tau^*$ of the Gaussian function, as well as the aftershock duration $t_a$, 
%by minimizing the mean-square residuals of the 
%logarithm of the seismicity rate (using a simplex algorithm). 
using a maximum likelihood approach. 
We maximize the log-likelihood function defined by 
\be
L= \sum_{i=1,N} \log R(t_i) - \int  \limits_{t_{\rm min}}^{t_{\rm max}} R(t) dt \,,
\label{LL}
\ee
where the seismicity rate is given by 
\be
R(t)=  \int  \limits_{-\infty}^{\infty} R(t,\tau) \, {e^{ - (\tau+\sigma_0)^2/2{\tau^*}^2} 
\over  \tau^* \sqrt{2\pi}} \, d\tau \,.
\label{RG}
\ee
%bs added discussion on what log-likelikhood weighting is:
The log-likelihood function is maximized when the rate estimate $R$,
weighted logarithmically, is large when events occur at times
$t_i$, and when the total rate estimate integrated over time is low.

Table~1 gives the parameters of each simulation, and the results of the inversion.
Figures~\ref{figpdfs1} and \ref{figpdfs3} 
show the real stress distribution (evaluated from the stress map
shown in Figure~\ref{figUS}b) and the inverted one, for each 
synthetic aftershock catalog. We test both inversion methods, either 
solving (\ref{alpha}) for $P_{\tau}(\tau)$ for $ 80 <\tau < 80$ MPa, 
or assuming a Gaussian stress distribution.

Figure~\ref{figRt1} compares the theoretical seismicity rate given by (\ref{Rt1}) 
using the observed stress field, with the seismicity rate estimated 
from the seismicity catalog using (\ref{kern}), and with 
the reconstructed seismicity rate estimated using 
 (\ref{Rt2}) from the inverted stress distribution.  
For this synthetic catalog, the seismicity rate is almost indistinguishable 
from an Omori law with an exponent $p=0.93$ for $t/t_a<10$.

In the first 2 catalogs in Table~1, with more than several thousands events, we obtain a 
very good estimation (error less than 6\%) on all parameters $\tau^*$, $\sigma_0$ and $t_a$.
If the number of events decreases to 392 events, without changing the time interval, we still obtain a rather good estimation of $t_a$ and $\tau^*$, but the error on the stress drop increases (see model 3 in Table~1). 
For a shorter catalog (\#4 in Table~1), with 292 events and only 4 ranges of magnitude in time, 
the stress drop is not constrained, unless we fix the aftershock duration to its true value.
Alternatively, we can fix the stress drop and obtain a rather good estimation of $\tau^*$ and $t_a$.
This shows that the main effect in recovering $\sigma_0$, $\tau^*$ and $t_a$ is the 
catalogue time interval, which needs to extend over a reasonable fraction of $t_a$.
This is because very different values of $t_a$ and $\sigma_0$ can
produce very similar seismicity rate $R(t)$ for $t<t_a/100$, as can be shown in
Figure~\ref{figRt3}.
%bs modified line below
If we decrease $A\sigma_n$, keeping $\tau^*$ fixed, the
Omori exponent becomes  closer to 1, and the error on 
all parameters increases (see models \#5 and \#6).

When inverting for the complete distribution $P_{\tau}(\tau)$, 
the results are pretty good for the first simulation, with an 
unrealistic large time interval and number of events. 
%bs QQ: regarding the next sentence, 
% how do you get a value outside where you inverted when doing
% the complete distribution, which is not constrained to be gaussian?
There are however deviations in the tails, for $\tau > 30$ MPa, 
which correspond to very short corner times $ c = 
 t_a \, \exp(-\tau/A\sigma_n)= 10^{-13}$, much smaller than the minimum time 
$t_{\rm min} / t_a = 10^{-10}$ used for the inversion of $P_{\tau}(\tau)$.   
For catalogs \#2-4 in Table~1, 
the distribution of $P_{\tau}(\tau)$ is not constrained for $\tau<0$, 
and for $\tau \gg 1$, because of the limited time interval.    
The results are very poor for both simulations \#5 and \#6 in Table~1, 
with $A\sigma_n=0.1$ MPa and Omori exponent $p=0.993$. In this case, 
we have almost no resolution on $P_{\tau}(\tau)$ for $\tau<0$. This method only 
provides a rough estimate of the width of the distribution for $\tau>0$.
%bs added sentence below:
%Thus, in practice, unless one has a very long catalogue in time, and $p$
%significantly different from 1, little can be said about the stress
%shadow regions.
% ah: modified bellow, the information about stress shadow does not come
% from the time interval when we observe Omori law, ie from p, but for larger times 
Thus, in practice, unless one has a very long catalogue in time, and significant deviations from Omori law, little can be said about the stress shadow regions.

\subsection{ Off-fault aftershocks}

We can make simple estimates of the stress change and seismicity 
rate off of the fault plane. 
For mode III rupture, static elasticity reduces to a Laplacian $\triangle u=0$. 
For a Laplacian, a Fourier mode with
wavenumber $k$ along an infinite fault decays exponentially
into the bulk proportional to $k$ times the distance $y$ to the fault.
With these basis functions, we can easily extrapolate off of the fault,
although since it neglects rupture end effects, it is valid only for
distances less than the rupture length $L$
and in areas along-side the mainshock rupture area, and not extending
into the lobes of increasing stress beyond the finite rupture length.
Thus, we are looking at regions which would be in the
stress "shadow" of a simple rupture.
Within this region,
at a distance $y<L$ from the fault, the power-spectrum of the displacement 
for the modified slip model (\ref{uk23}) becomes 
$u(k,y) \sim (kL +1)^{-n} ~ \exp(-ky)$.
The power-spectrum of the stress change is given by
\be
\tau(k,y) \sim { k \, u(k,y) } \sim  { e^{-k y} \over (kL + 1) ^{n-1} }  \,.
\label{taukd}
\ee
This shows how the stress heterogeneity decays very rapidly with distance from the
rupture surface. 

\begin{center}
\begin{figure}
\includegraphics[width=0.85\textwidth]{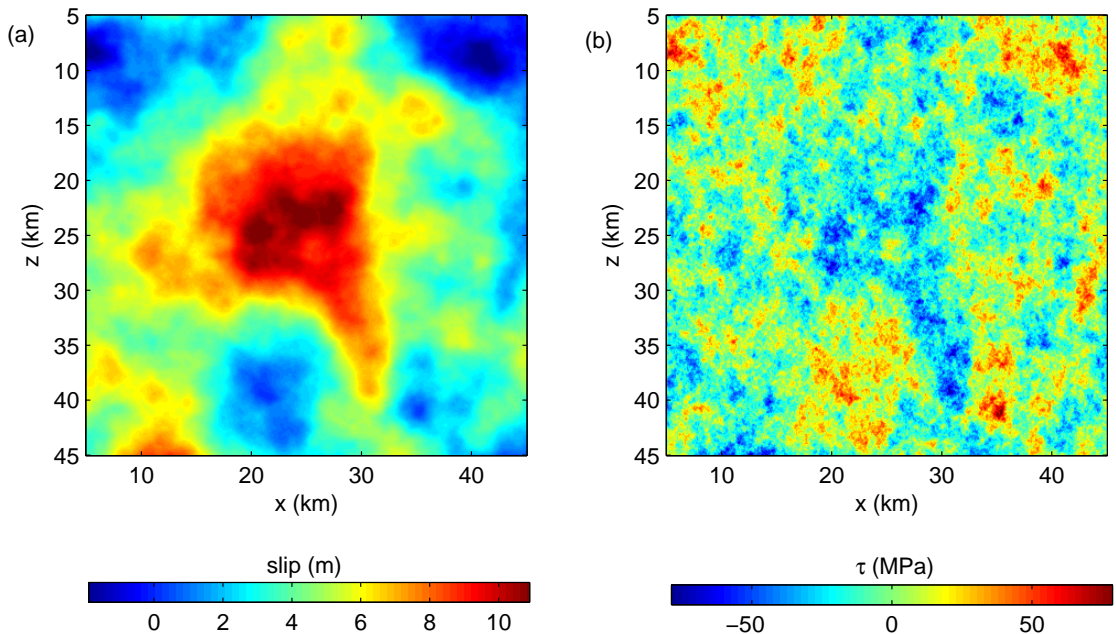} 
\caption{\label{figUS} (a) Stochastic slip model, with a power-spectrum 
$u(k) = 1 / (k L + 1)^{2.3}$ (where $L=50$ km is the rupture length), 
a stress drop $\sigma_0=3$ MPa, and a cell size $dx=0.1$ km. 
(b) Shear stress change (parallel to the slip direction).}
\end{figure}
\end{center}

\begin{center}
\begin{figure}
\includegraphics[width=0.5\textwidth]{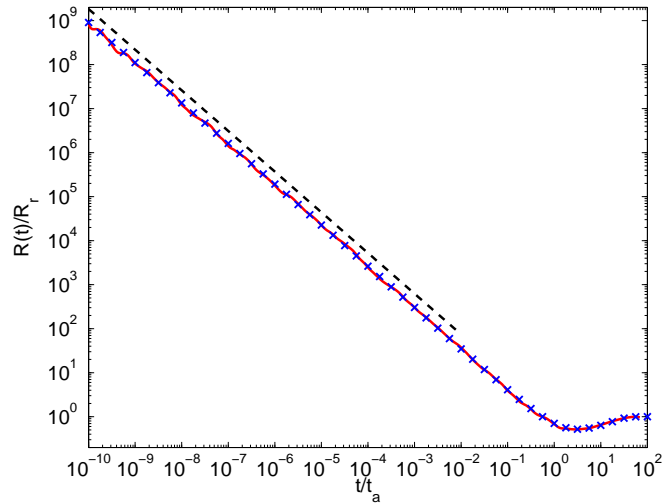}
% figUSR.m
\caption{\label{figRt1} 
Seismicity rate given by the rate-and-state model [{\it Dieterich}, 1994],
for the stress change shown in Figure~\ref{figUS}, assuming $A\sigma_n=1$ MPa, 
and without earthquake interactions. 
The solid red line is the seismicity rate estimated from the simulated earthquake catalog 
(see model \#1 in Table~1).
The dashed black line is a fit by Omori's law for $t <t_a/100$, 
with exponent $p=0.93$.
The crosses show the fit with the rate-and-state model assuming a Gaussian $P_{\tau}(\tau)$. } 
\end{figure}
\end{center}

\begin{center}
\begin{figure}
\includegraphics[width=0.5\textwidth]{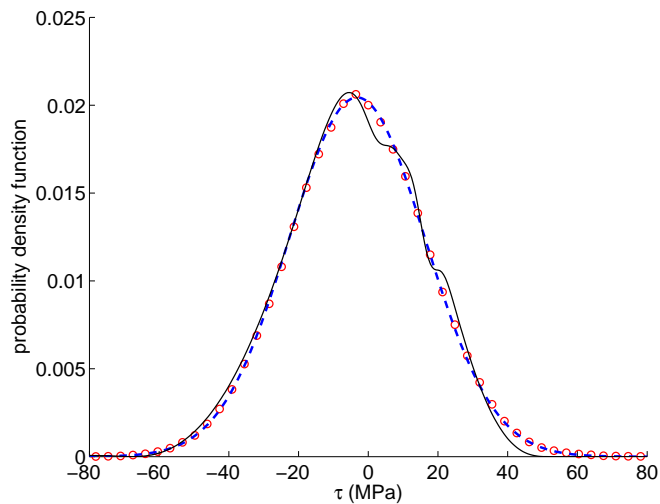}
\caption{\label{figpdfs1} 
Stress distribution estimated directly from the stress map 
shown in Figure~\ref{figUS}b (circles), 
and inverted from the seismicity rate shown in Figure~\ref{figRt1}. 
The solid black line is the solution of equation (\ref{alpha})). 
The best fitting Gaussian distribution is shown as a blue dashed line. } 
\end{figure}
\end{center}

\begin{center}
\begin{figure}
\includegraphics[width=0.5\textwidth]{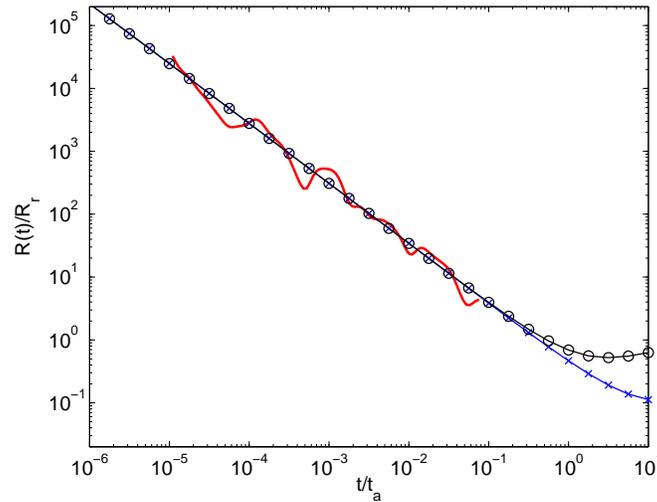}
\caption{\label{figRt3}
Seismicity rate given by the rate-and-state model [{\it Dieterich}, 1994],
for the stress change shown in Figure~\ref{figUS}, assuming $A\sigma_n=1$ MPa, 
and without earthquake interactions. 
The solid red line is the seismicity rate estimated from the simulated earthquake catalog 
(see catalog \#4 in Table~1).
The crosses show the fit with the rate-and-state model assuming a Gaussian $P_{\tau}(\tau)$, 
and inverting for $t_a$, $\sigma_0$, and $\tau^*$. 
The circles are the fit assuming a Gaussian $P_{\tau}(\tau)$ 
with the stress drop fixed to its real value.} 
\end{figure}
\end{center}

\begin{center}
\begin{figure}
\includegraphics[width=0.5\textwidth]{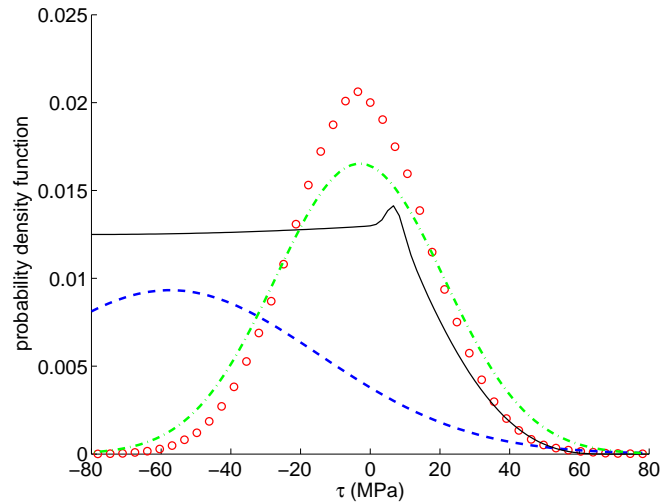}
\caption{\label{figpdfs3}
Stress distribution estimated directly from the stress map 
shown in Figure~\ref{figUS}b (circles), 
and inverted from the seismicity rate shown in Figure~\ref{figRt3} (see model \#4 in Table~1). 
The solid black line is the solution of equation (\ref{alpha})). 
The best fitting unconstrained Gaussian distribution is shown as a blue dashed line. 
The green dash-dot line shows the best fitting Gaussian with fixed stress drop.}
\end{figure}
\end{center}

\begin{figure}
% see code Rtdist.m
\includegraphics[width=1\textwidth]{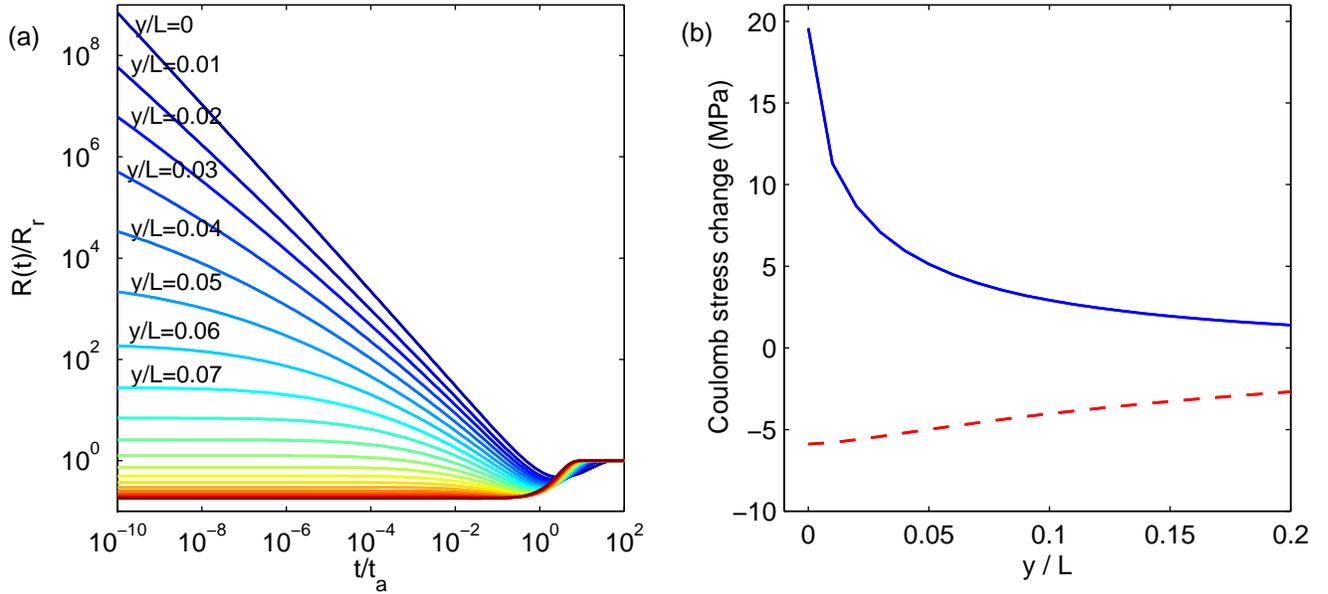}
\caption{\label{figRd} 
(a) Seismicity rate for different values of the distance to the fault $y/L$ decreasing 
from $y/L=0$ (top) to $y/L=0.2$ (bottom), 
using the slip model shown in Figure~\ref{figUS}a, and assuming $A\sigma_n=1$ MPa. 
(b) Standard deviation (solid line) and absolute value of the mean (dashed line, the average stress change is always negative) of the stress distribution as a function of the distance to the fault.}
\end{figure}

Figure~\ref{figRd}a shows the seismicity rate for different values of the 
distance from the fault $y/L$, using the slip model shown in Figure~\ref{figUS}a.
We computed the stress at a distance $y$ from the fault using 
$\tau(k,y) = \tau(k,0) \, \exp(-k \, y)$, i.e., multiplying the stress 
map shown in Figure~\ref{figUS}b by $\exp(-k \, y)$ in the Fourier domain.
The stress distribution is reasonably close to a Gaussian distribution at all distances. 
Therefore, we have used the best-fitting Gaussian distribution in order to compute the seismicity rate shown in Figure~\ref{figRd}a.
The standard deviation of the stress distribution decreases very fast with the distance 
to the fault, which produces a strong drop of the seismicity rate off of the fault. 
The average stress decreases much slower with $y$. 
%bs added sentence below:
Figure~\ref{figRd}b shows the falloff with distance of these quantities.
For $y/L>0.1$, the stress field is 
much more homogeneous and mostly negative (the standard deviation is smaller than 
the absolute mean stress). 
Therefore, the seismicity rate for $y/L>0.1$ is smaller 
than the reference rate at all times $t<t_a$.
 {\it Marsan} [2006]  also used the rate-and-state model to investigate 
how stress change heterogeneity modifies the rate of off-fault aftershocks.
He used a slightly different slip model, and 
%bs replaced with the sentences below:
assumed the spatial dependence of the stress variability decayed with the same
form as the stress, as distance cubed.  With this assumption he found,
not surprisingly, larger distances of triggering.

In practice, it is difficult to analyze the rate of off-fault aftershocks,
because the aftershock rate and the reference seismicity rate decrease 
with the distance from the fault, and because of location errors.
Also, secondary aftershocks triggered 
by off-fault events will perturb the stress field and seismicity rate 
with additional stress heterogeneity. 
Our seismicity rate estimates here presume focal mechanisms of aftershocks
similar to the mainshock focal mechanism; other focal mechanisms
could have different rates, but optimally oriented plane estimates
may not be the best approach [{\it McCloskey et al.}, 2003].  In any case,
we do see very rapid falloff of the seismicity with distance from the fault, 
a point which deserves further observational exploration.
%AA
Note that Figure~\ref{figRd}a shows the seismicity rate normalized by the reference 
rate $R_r$. If $R_r$ decreases with the distance to the fault, 
the decrease of the aftershock rate with $y$  will be even faster than 
shown in Figure~\ref{figRd}a.

\subsection{ Gaussian versus exponential stress distribution}

While the pure Omori law with $p<1$ occurs for
the exponential distribution of  stress changes, we find
numerically that a Gaussian stress distribution
(which the $k^2$ model and many other models give),
also gives realistic looking $p$-values over wide ranges of time scales.
Some insight into why this is the case can be gained by noting that
for a sufficiently wide range of values, a Gaussian is a good enough
approximation of an exponential.  Taking the ratio of a Gaussian
to an exponential
\ba
 &&\exp \left(- { (\tau + \sigma_0)^2 \over 2 {\tau^*}^2} \right) \,
  /\exp \left( {- \tau \over \tau_0} \right) = \nonumber \\ 
&&\exp \left[ - {1 \over 2} \left({\tau +\sigma_0 \over  {\tau^*}} 
- {{\tau^*} \over \tau_0} \right)^2-{\sigma_0\over \tau_0}
+{{  {\tau^*}}^2 \over 2 {\tau_0}^2}  \right] \, . 
\label{expgau}
\ea
For ${\tau /  {\tau^*}}={ {\tau^*} / \tau_0}
-{\sigma_0 /  {\tau^*}} \pm 1$
this is within a factor  $\exp(1)$ of being constant. %%
Thus, over an $e$-folding range of ${\tau^*}/\tau_0$ we have something well
approximated by an exponential.

\begin{center}
\begin{figure}
% see code pgauss.m
\includegraphics[width=0.5\textwidth]{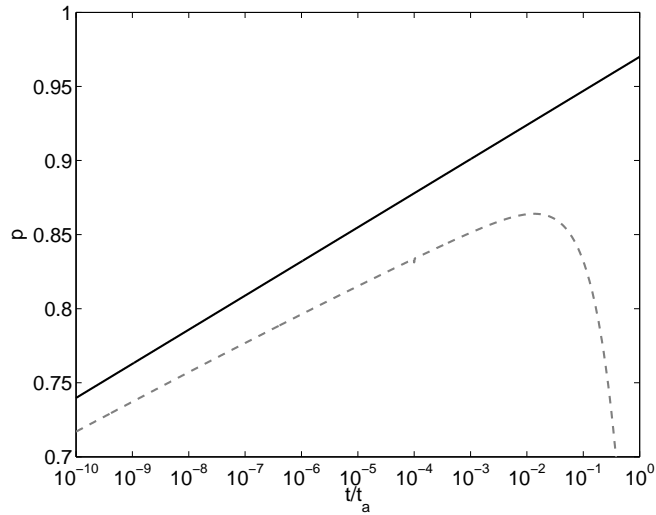}
\caption{\label{figpt} 
Variation of the effective Omori exponent with time, for a Gaussian stress distribution 
of mean $-\sigma_0=-3$ MPa and standard deviation $\tau^*=10$ MPa. 
The dashed line is the exact solution (given by integrating numerically 
(\ref{Rt2}), and the solid line is the 
approximate analytical solution (\ref{pt}).}
\end{figure}
\end{center}

We can use this result to obtain an approximate analytical expression for the effective 
Omori exponent produced by a Gaussian stress distribution.
Expression (\ref{expgau}) shows that the exponential distribution
closer to the Gaussian one for a stress $\tau$ has a characteristic parameter 
$
\tau_0 =  {\tau^*}^2 /(\sigma_0 + \tau) \,.
$
As Figure~\ref{figRtPexp} illustrates,
the more important contribution to the aftershock rate at time $t\ll t_a$ 
is due to stress values of the order of $\tau_c = -  A\sigma_n \log(t/t_a)$.
 If the stress change obeys a Gaussian distribution, 
stresses larger than $\tau_c $ are less frequent than for $\tau= \tau_c$, 
therefore they have a smaller contribution to the seismicity rate  at time $t$. 
Smaller stress values $\tau < \tau_c $ are more frequent, but the seismicity 
rate at time $t$ is negligible compared to larger stress  values. 
 We thus obtain the following relation between the parameters $\sigma_0$ and 
 $\tau^*$ of a Gaussian distribution, and the parameter $\tau_0$ of the exponential 
 distribution which better explains the aftershock rate at a given time $t$
 \be
\tau_0 =  { {\tau^*}^2 \over  \sigma_0 -  A\sigma_n \log(t/t_a)  }
 \ee
  Using expression (\ref{p}), we obtain the following relation between  the effective 
  Omori exponent at time $t$ and the parameters  $\sigma_0$ and 
  ${\tau^*}$ of a Gaussian stress distribution 
\be
p \approx 1 - { A \sigma_n \sigma_0 -  A^2 \sigma_n^2 \log(t/t_a) \over { {\tau^*} } ^2} \,
\label{pt}
\ee  
showing the slow increase of $p$ with time.
Figure~\ref{figpt} compares this approximate solution (\ref{pt}) with 
the variation of $p$ with time computed by integrating numerically (\ref{Rt2}), 
using $A\sigma_n=1$ MPa, for a Gaussian stress distribution with $\sigma_0=3$ 
MPa and ${\tau^*} =10$ MPa.
The approximate solution (\ref{pt}) for Omori exponent is 
%bs replaced with word below: very 
quite 
good for
short times $t\ll t_a$, but the difference with the exact solution increases 
as time approaches $t_a$.
This expression  (\ref{pt})  also shows us the inherent tradeoff between 
the mean stress change $-\sigma_0$ and the variance of
the stress change ${\tau^*}$ in affecting the $p$-value.

\section{ Discussion}
%%%
                                                                                                                                                                                                                                                                                                                                                                                                                                                                                                                                                                                                                                   We have considered above only heterogeneity of the Coulomb stress change.
However, there are other important factors that affect the temporal evolution 
of the seismicity rate, such  as heterogeneity of the friction law parameter $A$, normal stress, 
and stressing rate, multiple interactions between aftershocks, foreshocks, and postseismic relaxation.

\subsection{ Heterogeneity of the friction parameter $A$, normal stress, stressing rate, 
and reference seismicity rate.}

We have shown that Coulomb stress change heteroegeneity modifies  the 
temporal evolution of the seismicity rate, compared to a uniform stress change.  
But other kinds of heterogeneity may also impact the aftershock decay with time, 
in particular the normal stress.
Normal stress heterogeneity enters the problem in two ways, through the
``modified'' Coulomb stress change $\tau$, and through the $A\sigma_n$ term in 
the denominator.  

Slip on a rough fault will produce coseismic changes of the normal stress [{\it Dieterich}, 2005].
For coseismic changes of the normal stress  which are small compared to the normal stress, 
we can assume that $A\sigma_n$ does not change with time,  and account for coseismic  
changes of $\sigma_n$ only in the coseismic Coulomb stress change $\tau$. 
For larger coseismic changes of normal stress, we have to use a more complex form for the relation 
(\ref{dg}) between stress history and seismicity rate [{\it Dieterich}, 1994], and equation
 (\ref{R}) is no more valid.

In addition to coseismic stress changes of $\sigma_n$, there are also  
spatial fluctuations of $A\sigma_n$. For instance, we expect both $A$ and 
$\sigma_n$ to change with depth. 
With a wide variety of
materials making up fault zones and the presence of fluids, there 
%bs replaced: are 
is 
probably
no  lower bound on $A \sigma_n$.
The first effect of introducing heterogeneity of $A\sigma_n$ is to increase the fluctuations of the 
normalized stress $\tau/A\sigma_n$, i.e., the standard deviation of  $\tau/A\sigma_n$ 
is larger than $\tau^*/\langle A\sigma_n \rangle$. Therefore, Omori exponent increases if
$A\sigma_n$ is more heterogeneous.
Neglecting the fluctuations of $A\sigma_n$ will thus overestimate   $\tau^*$.
 
 The second effect is to introduce fluctuations of the aftershock duration $t_a$, which 
 scales with $A\sigma_n$. Fluctuation of $A\sigma_n$ thus decrease the apparent 
 aftershock duration. The value of $t_a$, inverted assuming $\tau/A\sigma_n$ is uniform, 
 is smaller than its average value $t_a$. Also,  $A\sigma_n$ heterogeneity 
 widens the duration of the quiescence (time period when $R(t) <R_r$).
 
\begin{center}
\begin{figure}
\includegraphics[width=0.5\textwidth]{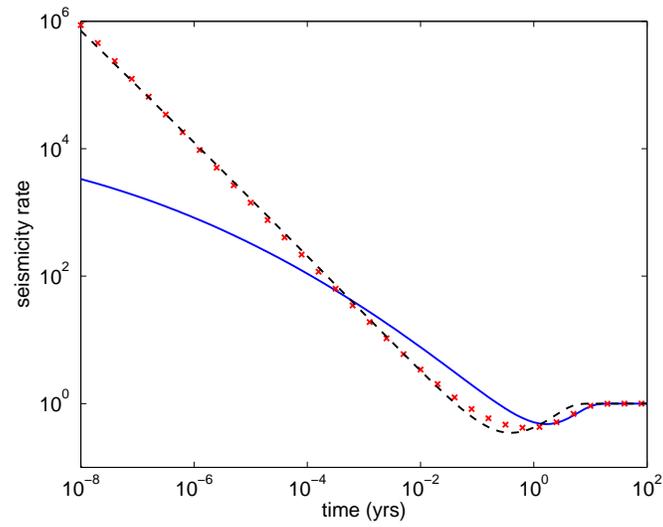}
% see code RS-PS/ptauoas.m	and mat output file resptauoas.mat
\caption{\label{figRtPtauPas} 
Seismicity rate for a Gaussian stress distribution ($\sigma_0=-3$ MPa, $\tau^*=5$ MPa), 
without (solid line) and with (crosses) heterogeneity of $A\sigma_n$. Normal stress fluctuations 
are modeled by a log-normal distribution of average 1 MPa and standard deviation 7.3 MPa.
The dashed line is a fit by the rate-and-state model (\ref{R}), assuming  $A\sigma_n$ is uniform.}
\end{figure}
\end{center}

 We have illustrated the effect of normal stress heterogeneity in Figure \ref{figRtPtauPas}, 
 which compares the   seismicity rate with and without fluctuations of $A\sigma_n$.
 Fluctuations of coseismic Coulomb stress changes are modeled by a Gaussian 
 distribution 
of mean $-\sigma_0=-3$ MPa and standard deviation $\tau^*=5$ MPa. 
%bs QQ: Fig 12 caption says std=4Mpa; which is it??
%ah:   $std(A\sigma_n)=7.3$ MPa., i've corrected the caption
 For  $A\sigma_n$, we use a lognormal distribution of mean $ \langle A\sigma_n \rangle=1$ MPa 
 and standard deviation $std(A\sigma_n)=7.3$ MPa.
The main effects of $A\sigma_n$ heterogeneity is to increase the apparent Omori exponent 
(measured for $t<0.01$ yr)  from 0.44 to 0.91, and to decrease the apparent aftershock duration, 
(defined as the time when the aftershock rate decreases below its reference value) 
from 0.2 to 0.05 yr. 

Inverting for the Coulomb stress distribution from $R(t)$, assuming that 
$A\sigma_n=1$ MPa everywhere, gives $\tau_g^*=15.7  $ MPa, $\sigma_{0,g}=9.9$ MPa and $t_{a,g}=0.15$ yr, 
instead of the true value   $\tau^*=5$ MPa, $\sigma_{0,g}=3$ MPa and $\langle t_{a} \rangle=1$ yr.
% see matlab code ptquoas.m. and matlab result file resptauoas.mat / varable: RES3
The errors in the inverted parameters  $\tau_g^*$, $\sigma_{0,g}$, $t_{a,g}$ are negligeable when 
 the Coulomb stress change $\tau$ is more heterogeneous than $A\sigma_n$, i.e., if 
 $\tau^* \gg std(A\sigma_n)$. 
The fit of the aftershock rate with $A\sigma_n$ assumed constant gives a reasonably good fit 
to the seismicity rate computed including  $A\sigma_n$ heterogeneity. 
The misfit will probably be within the noise level for real data set. 
This shows that, with the time dependence of the seismicity alone being the source of
information, we cannot distinguish between  heterogeneity of $\tau$ or $A\sigma_n$.
Finding other effects which might be able
to separate out these contributions of shear stress heterogeneity
and normal stress heterogeneity remains an area worthy of further inquiry.

The fact that  $A\sigma_n$ heterogeneity increases the Omori exponent may explain why 
very low $p$-values are seldom observed, even outside the mainshock rupture area, where 
Coulomb stress change is relatively uniform (see Figure (\ref{figRd}b).
This also explains why the crossover time $c$ for off-fault aftershocks is usually very short, 
 as for on-fault aftershocks.
It also makes  stress shadows even more difficult to detect. 
Even in the regions where stress change is negative 
and not too heterogeneous ($\tau^* < \sigma_0$), fluctuations of $A\sigma_n$ produce 
an increase of the seismicity rate at short times, while a uniform value of $A\sigma_n$ 
gives a quiescence at all times.

Another parameter that affects the seismicity rate is the reference stressing rate, which  modifies 
the aftershock duration $t_a$. Heterogeneity of the stressing rate will thus also yield an error in  
the inverted values of $\tau^*$ and  $\sigma_{0}$. 
In contrast, the seismicity rate does not depend on the spatial fluctuations of the reference 
seismicity rate, but depends only on the average value of $R_r$. 
In practice, $R_r$ is measured from the average seismicity rate over 
a long time period before the mainshock. The uncertainty on $R_r$ is generally of a factor of about 2.
This could induce large relative errors on the  stress drop estimate $\sigma_{0,g}$, but 
does not affect too much the inverted values of $\tau^*$ and $t_a$.

%If both $tau$ and $A\sigma_n$ are heterogeneous, the seismicity rate is given by 
%\be
%R(t)  =  \int \limits_{-\infty}^{\infty} \, R(t,\tau/A\sigma_n) P_{\tau/A\sigma_n}  d(\tau/A\sigma_n) 
%\label{Rtauas}
%\ee
%We can thus apply the same methods to invert for $P_{\tau/A\sigma_n}$ from the seismicity rate $R(t)$.

%If fluctuations of $\tau$ and $A\sigma_n$ are decoupled, then the distribution of 
%$\tau/A\sigma_n$ is given by
%\be
%P_{\tau/A\sigma_n} (u) =  \int \limits_0^{\infty} \, A\sigma_n 
%P_{A\sigma_n} (A\sigma_n) \, P_(\tau} (u A\sigma_n) \, d(A\sigma_n)
%\ee
%If $A\sigma_n$ is very heterogeneous (standard deviation much larger than the average value), 
%the distribution $P_{\tau/A\sigma_n}$ will have a different shape from $P_{\tau}$. The effect
%of introducing fluctuations of  $A\sigma_n$ is to increase the heterogeneity of $\tau/A\sigma_n$, 
%and thus to increase Omori exponent.
%Assuming that $A\sigma_n$ is constant will thus over-estimate the coseismic stress heterogeneity 
%$\tau^*$. For instance, if  $A\sigma_n$ has a lognormal distribution with a mean value of 1, 

\subsection{ Foreshocks}
An assumption of our model is that the seismicity rate before the mainshock is equal 
to the reference seismicity rate.
But most mainshocks are preceded by foreshocks, so that 
the seismicity rate $R_0$ before the mainshock is usually larger than
the reference rate $R_r$. Using the results of {\it Dieterich} [1994], 
we can take into account this effect by replacing 
the term $e^{-\tau/A\sigma_n}$ in (\ref{R}) by 
\be
 {R_r \over R_0 } \exp \left( - {\tau \over A\sigma_n } \right) = 
 \exp \left(- {\tau \over A\sigma_n }  - \log \left( { R_0 \over R_r }  \right)  \right) \,.
\ee 
The effect of increasing $R_0$ is thus equivalent to shifting the stress distribution  
toward larger values, by the amount $A\sigma_n \, \log( R_0/R_r)$.
Not correcting for this effect will thus over-estimate the stress change.

\subsection{Secondary aftershocks}
We have neglected in this study the role of aftershocks in changing
the seismicity rate and redistributing the stress.
We know that most aftershocks may be secondary aftershocks, triggered 
by previous aftershocks [{\it Felzer et al.}, 2003; {\it Helmstetter and 
Sornette}, 2003]. 
{\it Ziv and Rubin} [2003] studied a quasi-static fault model that is 
governed by rate- and state-dependent friction.
They have shown that, if the mainshock is modeled as 
a uniform stress increase, the effect main of secondary aftershocks in the 
rate-and-state model is to renormalize the 
seismicity rate without changing its time dependence (i.e., without changing Omori $p$ value). 
If the stress change induced by the mainshock is non-uniform, multiple 
interactions between earthquakes modify the spatial distribution of 
aftershocks [{\it Ziv}, 2003].

%% ref added
{\it Marsan} [2006]  also performed numerical simulations to model the effect of multiple interactions. 
He modeled the stress change induced by each aftershock by a Gaussian white noise of zero mean, 
i.e., assuming  all aftershocks have the same size, and neglecting spatial correlation of the stress field.
He concluded that the main effect of multiple interactions is  to increase the reference rate, but also 
to decrease the ratio of the aftershock and background rates. 
The existence of  multiple interactions also decreases the apparent aftershock duration, 
but does not change the Omori exponent.

Therefore, secondary aftershocks should not change the value of the width 
of the stress distribution inverted from the aftershock decay on the mainshock fault, which 
is controlled by Omori exponent.
But multiple interactions may bias the value of the average stress change and aftershock duration. 
Developing more realistic models for multiple interactions remains an area worthy of further inquiry,
but beyond the goals of this paper.

\section{Conclusion}
We have shown how a new extension of the rate-and-state friction formulation
for seismicity rates, which takes as its foundation a heterogeneous stress field, 
can explain the most prevalent and puzzling of aftershocks,
those which occur on the mainshock rupture area, where the stress decreases 
on average after the mainshock.  With this point of view,
subtle but significant deviations from the pure inverse time omori
exponent are mapped onto measures of stress change heterogeneity on the fault.
This contrasts with the established methodology of 
{\it Dieterich et al.} [2000, 2003], in which these deviations are mapped
onto time dependent stress changes following the mainshock.

Taking the point of view that stress heterogeneity can be quite
large at the local scale on the fault surface which ruptured,
we have gained insights into a number of topics of relevance
to stress heterogeneity's and earthquake behavior.
Regarding stress shadows, we have seen how they are very difficult
to detect in a heterogeneous stress context, relying on subtle
details in the seismicity rates at times of order $t_a$,
subtleties which would become even more difficult to detect if
$t_a$ were nonuniform.
%BS but what about at large distances like Stein et al?

Regarding stress heterogeneity relative to mean stress changes,
we have found in our efforts to model seismicity changes with
scale invariant slip distributions that typical stress changes
are actually larger than mean stress drops on faults, so that
a picture of a very rough stress distribution on a fault
which has broken is a much better picture than standard 
crack-like models, which tend to concentrate aftershocks
at the edges of ruptures.  
This provides important constraints on physical models of earthquakes.
Finally, we have shown that modest catalogue lengths allow an accurate
inversion for some stress heterogeneity parameters, if the only source of
 heterogeneity is the  Coulomb stress change.
%Since stress heterogeneity is a huge topic of relevance to a wide
%variety of seismological phenomena, application of this technique
%to real catalogues, which we are currently pursuing,
%should be very exciting.

%%
However, there are probably other important factors that affect the temporal evolution 
of the seismicity rate, such  as heterogeneity of the friction law parameter $A$, effective normal stress, 
and stressing rate, multiple interactions between aftershocks, and postseismic relaxation.
In particular, heterogeneity of  $A\sigma_n$ may explain why Omori exponent and 
characteristic time $c$ does not seem to depend on stress change amplitude
%bs added reference below and in bibliography:
[{\it Felzer}, 2005].
We have shown that earthquake triggering is not only controlled by the average values of 
the Coulomb stress change, or of the effective normal stress,  but rather by their heterogeneity. 
Particularly, short time aftershock rate is mainly controlled by the maximum stress change in this region, 
rather than by its average value. 
Estimation of coseismic slip from seismograms or geodesy is not accurate enough to estimate 
small scale variations of the stress change on the fault plane. Therefore,  
we need to couple large scale deterministic slip models with small scale stochastic slip models, 
in order to reproduce the spatio-temporal distribution of triggered earthquakes.
This also shows the difficulty of inverting the stress field from the spatio-temporal variations 
of the seismicity rate. Real data is limited  in number of events, catalog duration, and 
location accuracy; and we have only rough estimates of the large-scale average value of  
the  friction parameters, normal stress and stressing rate. 
With the limited information given by the seismicity rate, 
it is hard to characterize the different factors that control earthquake triggering, especially 
on or close to the fault, where stress and material properties are very heterogeneous.

\begin{acknowledgments}
We thank Jim Dieterich, Fabrice Cotton, Michel Campillo and Alon Ziv 
for useful discussions. Jeffrey McGuire (associate editor) and Susanna Gross (reviewer) 
also provided interesting suggestions that helped improve the quality of the manuscript.
Part of this work was  done while the authors 
were at the KITP in Santa Barbara. 
This research was supported in part by the National Science Foundation under
grants  PHY99-0794 % KITP 
and EAR03-37226, % postdoc lamont
by the Southern California Earthquake Center (SCEC),
and by the Brinson Foundation. % postdoc lamont
% Add a SCEC grant #?
SCEC is funded by NSF Cooperative Agreement EAR-0106924 and
USGS Cooperative Agreement 02HQAG0008.
The SCEC contribution number for this paper is 930.

\end{acknowledgments}

{}

\end{article}

\clearpage

\begin{table}
\label{tab1}
% resinv_sim.txt
% figUSR.m
\caption{ Parameters of the synthetic aftershock catalogs: number $N$ of events,
time interval $[t_{\rm min} \, t_{\rm max}]$, value of $A\sigma_n$ used for the simulations, 
Omori exponent $p$ (measured by maximum likelihood from the simulated catalog for $t/t_a<0.01$), 
average stress change  $-\sigma_0$, and standard deviation $ \tau^*$ (in MPa), and results of the inversion: 
$\sigma_{0,g}$, $\tau^*_{g}$ and $t_{a,g}$, estimated
 assuming a Gaussian stress distribution $P_{\tau}(\tau)$.     Stress values are in MPa. }
 \begin{flushleft}
\begin{tabular}{|rrlrcc rcc ccl |}
\tableline
\#	 & $N$    & $t_{\rm min}$ & $t_{\rm max}$  & $p$ & $-\sigma_0$ & $A\sigma_n$ 
%bs modified \sigma_0 to \sigma_{0,g} in next line
& $-\sigma_{0,g}/A\sigma_n$  & $\tau*$ &    $\tau^*_{g}/A\sigma_n$ &  
$t_{a}$ & $t_{a,g}$  \\
\tableline
1 & 154447 & $10^{-10}$   & 100.  & 0.924 & -3.01 &   1.0 & -3.19& 
19.6 & 	19.5	&	$10^7$  & $1.00\times10^7$ 	\\ %k2.3_5.mat

2 &   3550 & $10^{-6}$ & 1.    & 0.938  &  -3.01 & 1.0   & -3.14  & 
19.6    & 20.6 & $10^7$  & $0.95\times10^7$   \\ %k2.3_6.mat

3\tablenotemark{a} &  392  & $10^{-6}$& 1.  &0.929 & -3.01 & 1.0 &-9.45&  
19.6  &20.8 & $10^7$  & $1.58\times10^7$  \\ %k2.3_62.mat
3\tablenotemark{a} &  392  & $10^{-6}$& 1.  &0.929 & -3.01 & 1.0 &-1.19&  
19.6  &18.3 & $10^7$  & ${1.00\times10^7}^\dagger$  \\ %k2.3_62.mat
3\tablenotemark{a} &  392  & $10^{-6}$& 1.  &0.929 & -3.01& 1.0 & $-3.01^\dagger$&  
19.6  &18.7 & $10^7$  & $1.12\times10^7$  \\ %k2.3_62.mat

4 &    231 & $10^{-5}$ &   0.1 & 0.948 &  -3.01 & 1.0 &-57.5 &
 19.6  & 42.8 & $10^7$  & $5.01\times10^7$  \\ %k2.3_7.mat
%bs modified 0. to 0.1 in next line 3rd column
4 &    231 & $10^{-5}$ &   0.1 &  0.948 &  -3.01 & 1.0 &$-3.01^\dagger$   &
 19.6  & 24.1 & $10^7$  & $0.90\times10^7$  \\ %k2.3_7.mat
4 &    231 & $10^{-5}$ &   0.1  & 0.948 &  -3.01 & 1.0 &-6.03 &
 19.6  & 25.8 & $10^7$  & ${1.00\times10^7}^\dagger$  \\ %k2.3_7.mat

 5 & 203998 & $10^{-10}$ &  100. & 0.995 & -3.01 & 0.1 & -30.8 &
 19.6 & 195. &  $10^7$  & ${1.00\times10^7}^\dagger$  \\ %k2.3_8.mat
 
6 &   3857 & $10^{-6}$ &    1.  & 0.992 & -3.01 & 0.1 & -45.9& 
19.6  & 133. & $10^7$  & $1.21 \times10^7$  \\ %k2.3_9.mat
  6 &   3857 & $10^{-6}$ &    1.  & 0.992 & -3.01 & 0.1 & -29.2& 
19.6  & 171. & $10^7$  & ${1.00 \times10^7}^\dagger$  \\ %k2.3_9.mat
6 &   3857 & $10^{-6}$ &    1. & 0.992 & -3.01  & 0.1& $-30.1^\dagger$& 
19.6  & 125. & $10^7$  & $1.09 \times10^7$  \\ %k2.3_9.mat

 \tableline
\end{tabular}
\tablenotetext{}{}		% to add a space
\tablenotetext{a}{This catalog is a subset of catalog \#2, obtained by increasing the 
minimum magnitude by one unit.}
%bs changed the tablenotetext from * to \dagger above and here so as
% not to confuse with the \tau^* notation:
\tablenotetext{\dagger}{The parameter was fixed to its real value in the inversion}

\end{flushleft}
\end{table}

\newpage

\begin{center}
\begin{figure}
\includegraphics[width=0.5\textwidth]{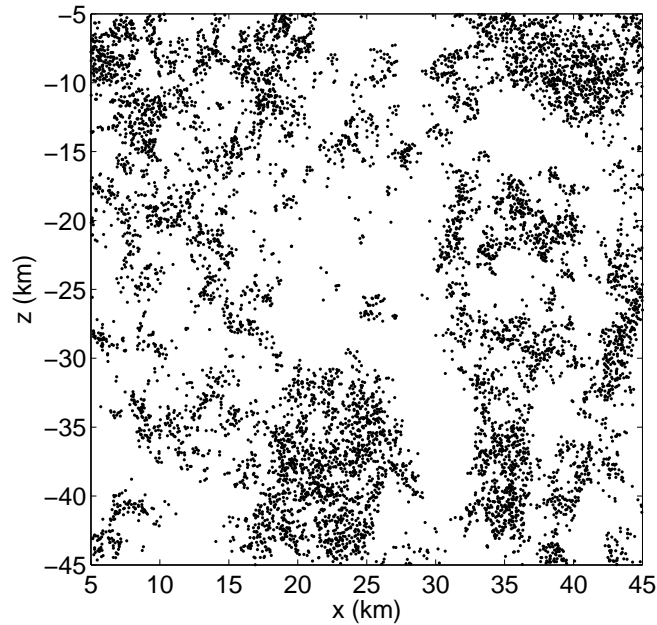}
\caption{\label{figmap} 
Seismicity map, for a synthetic catalog generated using the rate-and-state model
 [{\it Dieterich}, 1994],
for the stress change shown in Figure~\ref{figUS}, assuming $A\sigma_n=1$ MPa, 
and without earthquake interactions. Only events with $t<t_a$ are shown.}
\end{figure}
\end{center}

\end{document}